\documentclass[aps,prd,twocolumn,superscriptaddress,showpacs]{revtex4}
\usepackage[dvips]{graphicx}
\usepackage{amsmath,amssymb,times}
\usepackage{braket}

\newcommand{\bequ}{\begin{equation}}
\newcommand{\eequ}{\end{equation}}
\newcommand{\bea}{\begin{eqnarray}}
\newcommand{\eea}{\end{eqnarray}}

\def\gsim{~\,\makebox(1,1){$\stackrel{>}{\widetilde{}}$}\,~}
\def\lsim{~\,\makebox(1,1){$\stackrel{<}{\widetilde{}}$}\,~}

\DeclareSymbolFont{boldletters}{OML}{cmm} {b}{it}
\DeclareSymbolFontAlphabet{\mathbit}{boldletters}
\DeclareMathSymbol{\alpha}{\mathalpha}{letters}{"0B}
\DeclareMathSymbol{\beta}{\mathalpha}{letters}{"0C}
\DeclareMathSymbol{\gamma}{\mathalpha}{letters}{"0D}
\DeclareMathSymbol{\delta}{\mathalpha}{letters}{"0E}
\DeclareMathSymbol{\epsilon}{\mathalpha}{letters}{"0F}
\DeclareMathSymbol{\zeta}{\mathalpha}{letters}{"10}
\DeclareMathSymbol{\eta}{\mathalpha}{letters}{"11}
\DeclareMathSymbol{\theta}{\mathalpha}{letters}{"12}
\DeclareMathSymbol{\iota}{\mathalpha}{letters}{"13}
\DeclareMathSymbol{\kappa}{\mathalpha}{letters}{"14}
\DeclareMathSymbol{\lambda}{\mathalpha}{letters}{"15}
\DeclareMathSymbol{\mu}{\mathalpha}{letters}{"16}
\DeclareMathSymbol{\nu}{\mathalpha}{letters}{"17}
\DeclareMathSymbol{\xi}{\mathalpha}{letters}{"18}
\DeclareMathSymbol{\pi}{\mathalpha}{letters}{"19}
\DeclareMathSymbol{\rho}{\mathalpha}{letters}{"1A}
\DeclareMathSymbol{\sigma}{\mathalpha}{letters}{"1B}
\DeclareMathSymbol{\tau}{\mathalpha}{letters}{"1C}
\DeclareMathSymbol{\upsilon}{\mathalpha}{letters}{"1D}
\DeclareMathSymbol{\phi}{\mathalpha}{letters}{"1E}
\DeclareMathSymbol{\chi}{\mathalpha}{letters}{"1F}
\DeclareMathSymbol{\psi}{\mathalpha}{letters}{"20}
\DeclareMathSymbol{\omega}{\mathalpha}{letters}{"21}
\DeclareMathSymbol{\varepsilon}{\mathalpha}{letters}{"22}
\DeclareMathSymbol{\vartheta}{\mathalpha}{letters}{"23}
\DeclareMathSymbol{\varpi}{\mathalpha}{letters}{"24}
\DeclareMathSymbol{\varrho}{\mathalpha}{letters}{"25}
\DeclareMathSymbol{\varsigma}{\mathalpha}{letters}{"26}
\DeclareMathSymbol{\varphi}{\mathalpha}{letters}{"27}
\DeclareMathSymbol{\Gamma}{\mathalpha}{letters}{"00}
\DeclareMathSymbol{\Delta}{\mathalpha}{letters}{"01}
\DeclareMathSymbol{\Theta}{\mathalpha}{letters}{"02}
\DeclareMathSymbol{\Lambda}{\mathalpha}{letters}{"03}
\DeclareMathSymbol{\Xi}{\mathalpha}{letters}{"04}
\DeclareMathSymbol{\Pi}{\mathalpha}{letters}{"05}
\DeclareMathSymbol{\Sigma}{\mathalpha}{letters}{"06}
\DeclareMathSymbol{\Upsilon}{\mathalpha}{letters}{"07}
\DeclareMathSymbol{\Phi}{\mathalpha}{letters}{"08}
\DeclareMathSymbol{\Psi}{\mathalpha}{letters}{"09}
\DeclareMathSymbol{\Omega}{\mathalpha}{letters}{"0A}

\def\bk{\mbox{\boldmath $k$}}

\begin{document}
\title{
Determination of quark-hadron transition from lattice QCD and neutron-star observation
}

\author{Takahiro Sasaki}
\email[]{sasaki@phys.kyushu-u.ac.jp}
\affiliation{Department of Physics, Graduate School of Sciences, Kyushu University,
             Fukuoka 812-8581, Japan}

\author{Nobutoshi Yasutake}
\email[]{nobutoshi.yasutake@it-chiba.ac.jp}
\affiliation{Department of Physics, Chiba Institute of Technology,
             Chiba 275-0023, Japan}

\author{Michio Kohno}
\email[]{kohno@kyu-dent.ac.jp}
\affiliation{Physics Division, Kyushu Dental College,
             Kitakyushu 803-8580, Japan}

\author{Hiroaki Kouno}
\email[]{kounoh@cc.saga-u.ac.jp}
\affiliation{Department of Physics, Saga University,
             Saga 840-8502, Japan}

\author{Masanobu Yahiro}
\email[]{yahiro@phys.kyushu-u.ac.jp}
\affiliation{Department of Physics, Graduate School of Sciences, Kyushu University,
             Fukuoka 812-8581, Japan}

\date{\today}

\begin{abstract}
We determine the quark-hadron transition line 
in the whole region of temperature  ($T$) and baryon-number chemical potential ($\mu_{\rm B}$) from 
lattice QCD results and neutron-star mass measurements, making 
the quark-hadron hybrid model that is consistent with the 
two solid constraints. 
The quark part of the hybrid model is the Polyakov-loop extended Nambu-Jona-Lasinio (PNJL) model with entanglement vertex 
that reproduces lattice QCD results at $\mu_{\rm B}/T=0$, while 
the hadron part is the hadron resonance gas model with volume-exclusion effect that reproduces neutron-star mass measurements and the 
neutron-matter equation of state calculated from two- and three-nucleon forces based on the chiral effective field theory. 
The lower bound of the critical $\mu_{\rm B}$ of the quark-hadron transition at zero $T$ is $\mu_{\rm B}\sim 1.6$ GeV. 
The interplay between the heavy-ion collision physics around $\mu_{\rm B}/T =6$ and the neutron-star physics 
at $\mu_{\rm B}/T =\infty$ is discussed. 
\end{abstract}

\maketitle

\section{Introduction}
The phase diagram of quantum chromodynamics (QCD) is a key to understanding not only natural phenomena such as compact stars and the early Universe 
but also laboratory experiments such as relativistic heavy-ion collisions\cite{KL1994,Aoki,Fukushima-review}.
The first-principle lattice QCD (LQCD) simulation as a quantitative analysis of the phase diagram\cite{KL1994,Aoki}, 
however, has the severe sign problem at middle and large $\mu_{\rm B}/T$, 
where $T$ is temperature and $\mu_{\rm B}$ is baryon-number chemical potential. 
Therefore the QCD phase diagram is still unknown 
particularly at $\mu_{\rm B}/T \gsim 1$, although many possibilities are 
proposed by effective models there.
A steady way of approaching the middle and large $\mu_{\rm B}/T$ 
regions is gathering solid information from different regions and extracting a consistent picture from the information.

LQCD simulations are quite successful at $\mu_{\rm B}/T \lsim 1$\cite{KL1994,Aoki,AliKhan,Bazavov,Borsanyi}. 
They are providing high-precision results for the realistic 2+1 flavor system at the present day, 
for example the transition temperature, the equation of state (EoS), and fluctuations of conserved charges \cite{Bazavov,Borsanyi}.
As a way of extending the understanding to the $\mu_{\rm B}/T \gsim 1$ region, we can consider effective models such as the Polyakov-loop extended Nambu-Jona-Lasinio (PNJL) model \cite{Sakai-EPNJL,Sasaki-Columbia,PNJL8,Fukushima-Kashiwa,nlPNJL,Rossner,Sakai-extendedZ3,Gatto,Sakai-signproblem,Sakai-vector2008,Sakai-EOS,Meisinger,Fukushima,Megias,Ratti, Bratovic, Lourenco}.
Actually, some improved versions of the PNJL model yield 
desirable results consistent with LQCD simulations 
at $\mu_{\rm B}/T \lsim 1$~
\cite{Sakai-EPNJL, Sasaki-Columbia, nlPNJL, Fukushima-Kashiwa, PNJL8}. 
However, the model approach has still various ambiguity at large $\mu_{\rm B}/T$.

A key issue in the large $\mu_{\rm B}/T$ limit, i.e. at finite $\mu_{\rm B}$ but vanishing $T$, is the EoS of nuclear matter. 
It is one of the most important subjects in nuclear physics to understand properties of symmetric nuclear matter 
and neutron matter microscopically from realistic baryon-baryon interactions. 
Various theoretical frameworks have been developed to study the subject. 
The results seem to be reliable because most of them are now converging a common result, 
but the common result cannot reproduce empirical saturation properties properly 
if one starts with realistic two-nucleon forces (2NF).
This insufficiency is probably due to the lack of including three-nucleon forces (3NF). 
Recent development of the chiral effective field theory (Ch-EFT) \cite{WE79, CDB} 
provides a way of determining 2NF and 3NF systematically from 
symmetries of underlying QCD.
Although the Ch-EFT interaction is, by construction, to be applied at low and normal nuclear densities, 
the standard many-nucleon calculation using the Ch-EFT 2NF and 3NF at these densities 
should provide the predictive base for considering the neutron-matter 
EoS at higher densities. 
The combination of this new constraint and the experimental constraint~\cite{Danielewicz}
evaluated from the heavy-ion collision measurements is considered to be useful to determine the nuclear-matter EoS solidly. 

The mass-radius (MR) relation of neutron star (NS) is sensitive to the nuclear-matter EoS \cite{Lattimer}.
In this sense, astrophysical observation is another valuable source of information to provide a strong constraint on the EoS. 
Recent observations suggest the existence of massive NSs ($\sim 2 M_\odot $), which seems to exclude the possibility of soft EoS \cite{Demorest, Antoniadis}. 
However, there exists uncertainties on the radius of NSs from varying observations. 
Steiner {\it et al.} have adopted the statistical approach to constrain this uncertainty, and have provided the best fitting against various observations on the MR relation \cite{Steiner}. 

There is a possibility that  the quark-hadron phase transition occurs in NSs. 
The observations on the MR relation yield a strong constraint on 
both the quark and hadron phases, while the nuclear-matter EoS 
determined from the Ch-EFT 2NF and 3NF and the heavy-ion collision measurements does on the hadron phase. 
Therefore, the combination of the solid constraints  
may answer an important question, 
whether the quark-hadron phase transition occurs in NSs and 
further what is the critical chemical potential of the transition if 
it occurs. This is nothing but to clarify the QCD phase diagram 
in the large $\mu_{\rm B}/T$ limit. 

In this paper, we determine the QCD phase diagram in the whole region 
from $\mu_{\rm B}/T=0$ to infinity, constructing a reliable 
quark-hadron hybrid model. 
The quark part of the hybrid model is 
the Polyakov-loop extended Nambu-Jona-Lasinio (PNJL) model 
with entanglement vertex that reproduces LQCD data at 
finite imaginary $\mu_{\rm B}$, finite real- and imaginary-isospin chemical potentials, small real $\mu_{\rm B}$\cite{Sakai-EPNJL,Sasaki-Columbia}, and strong magnetic field\cite{Gatto}.
The hadron part of the hybrid model is the hadron resonance gas (HRG) model with volume-exclusion effect that reproduces 
the NS observations and the nuclear-matter EoS 
evaluated from the Ch-EFT 2NF and 3NF and the heavy-ion collision measurements. 
The volume-exclusion effect is necessary to reproduce the repulsive nature of the nuclear-matter EoS. 
The EoS provided by the hybrid model preserves the causality even at high $\mu_{\rm B}$.
In order to construct the nuclear-matter EoS from 
the Ch-EFT 2NF and 3NF, 
we employ the lowest-order Brueckner theory (LOBT) in pure neutron matter with the J\"{u}lich N$^3$LO interaction \cite{EHM09}. 
The lower bound of the critical $\mu_{\rm B}$ of the quark-hadron transition at $T=0$ is found to be 
$\mu_{\rm B}\sim 1.6$ GeV. 
We also investigate the interplay between the heavy-ion collision physics around $\mu_{\rm B}/T =6$ 
and the neutron-star physics at $\mu_{\rm B}/T =\infty$.

This paper is organized as follows. 
In Sec.~\ref{Modelsetting}, we present the quark-hadron hybrid model 
and evaluate the nuclear-matter EoS from the Ch-EFT 2NF and 3NF by using the LOBT. 
Numerical results are shown in Sec.~\ref{Numericalresults}. 
Section \ref{Summary} is devoted to a summary. 
\section{Model setting}
\label{Modelsetting}

We consider a two-phase model to treat the quark-hadron phase transition by assuming that the transition is the first order\cite{Shao,Blaschke,Bonanno:2011ch,Ippolito:2007hn}. 
For the quark phase, we use the entanglement PNJL (EPNJL) model \cite{Sakai-EPNJL,Sakai-EOS,Sasaki-Columbia}. 
This is an extension of the PNJL model and yields consistent results with LQCD data for finite imaginary $\mu_{\rm B}$, finite real- and imaginary-isospin chemical potentials, small real $\mu_{\rm B}$\cite{Sakai-EPNJL,Sasaki-Columbia}, and strong magnetic field\cite{Gatto}.
For the hadron phase, we use 
the HRG model. 
The model is successful in reproducing the QCD EoS below the transition temperature at $\mu_{\rm B}/T=0$ \cite{Bazavov,Borsanyi}.
This model is extended for the baryon part to include 
the volume-exclusion effect. 
The effect is necessary to reproduce the repulsive nature of the 
nuclear-matter EoS. The volume-exclusion radius is fitted 
to reproduce the nuclear-matter EoS determined 
from the Ch-EFT 2NF and 3NF and the heavy-ion collision measurements. 

In this work, we consider the 2-flavor system and 
do not take into account the existence of hyperons\cite{Schulze}. 
Even with hyperons, the fraction of hyperons is suppressed 
by the existence of quarks in NS~\cite{Maruyama2007}. 
Hence, the possibility of the appearance of quarks is first discussed 
in this paper. The possibility of the appearance of hyperons 
will be discussed in a forthcoming paper.  

\subsection{Quark phase}
We first consider the quark phase with the two-flavor EPNJL 
model. The Lagrangian density is obtained in Euclidean spacetime by 
\begin{eqnarray}
{\cal L}_{\rm EPNJL}
&=&
\bar{q}(\gamma_\nu D_\nu + {\hat m_0} - \gamma_4 {\hat \mu} )q
-G(\Phi )[(\bar{q}q )^2 +(\bar{q}i\gamma_5\vec{\tau}q )^2] 
\nonumber\\
&&
+{\cal U}(\Phi [A],\Phi^* [A],T) , 
\label{EPNJLmodel}
\end{eqnarray} 
where $D_\nu=\partial_\nu - i\delta_{\nu 4}A_{4}^{a}{\lambda_a /2}$ with the Gell-Mann matrices $\lambda_a$. 
The two-flavor quark fields $q=(q_{\rm u},q_{\rm d})$ have masses ${\hat m_0}={\rm diag}(m_{\rm u},m_{\rm d})$, 
and the quark-number chemical potential matrix ${\hat \mu}$ is defined by ${\hat \mu}={\rm diag}(\mu_{\rm u},\mu_{\rm d})$.
Baryon-number chemical potential is obtained by $\mu_{\rm B}=3(\mu_{\rm u}+\mu_{\rm d})/2$.

The gauge field $A_\mu$ is treated as a homogeneous and static background field.
The Polyakov-loop $\Phi$ and its conjugate $\Phi ^*$ are determined 
in the Euclidean space by
\begin{equation}
\Phi = \frac{1}{3}{\rm tr}_{\rm c}(L),~~~~~
\Phi^* = \frac{1}{3}{\rm tr}_{\rm c}({\bar L}),
\label{Polyakov}
\end{equation}
where $L  = \exp(i A_4/T)$ with $A_4/T={\rm diag}(\phi_r,\phi_g,\phi_b)$ in the Polyakov-gauge; 
note that $\lambda_a$ is traceless and hence $\phi_r+\phi_g+\phi_b=0$. 
Therefore we obtain 
\begin{eqnarray}
\Phi &=&{1\over{3}}(e^{i\phi_r}+e^{i\phi_g}+e^{i\phi_b})
\nonumber\\
&=&{1\over{3}}(e^{i\phi_r}+e^{i\phi_g}+e^{-i(\phi_r+\phi_g)}), 
\notag\\
\Phi^* &=&{1\over{3}}(e^{-i\phi_r}+e^{-i\phi_g}+e^{-i\phi_b})
\nonumber\\
&=&{1\over{3}}(e^{-i\phi_r}+e^{-i\phi_g}+e^{i(\phi_r+\phi_g)}) .
\label{Polyakov_explict}
\end{eqnarray}
We use the Polyakov-loop potential $\mathcal{U}$ of Ref.~\cite{Rossner}: 
\begin{eqnarray}
{\cal U}
&=&
T^4\Bigl[
-\frac{a(T)}{2} {\Phi}^*\Phi
\nonumber\\
&&
+ b(T)\ln(1 - 6{\Phi\Phi^*}  + 4(\Phi^3+{\Phi^*}^3) - 3(\Phi\Phi^*)^2 )
\Bigr]
\end{eqnarray}
with
\begin{equation}
a(T)
=
a_0
+ a_1\left(\frac{T_0}{T}\right)
+ a_2\left(\frac{T_0}{T}\right)^2,~~~~
b(T)
=
b_3\left(\frac{T_0}{T}\right)^3.
\end{equation}
The parameter set in $\mathcal{U}$ is fitted to LQCD data at finite $T$ in the pure gauge limit. 
The parameters except $T_0$ are summarized in Table \ref{table-para}.  
The Polyakov potential yields a first-order deconfinement phase transition at $T=T_0$ in the pure gauge theory. 
The original value of $T_0$ is $270$~MeV determined from the pure gauge LQCD data, 
but the EPNJL model with this value of $T_0$ yields a larger value of the pseudocritical temperature $T_\mathrm{c}$ of the deconfinement transition 
at zero chemical potential than $T_{\rm c}\approx 173\pm 8$~MeV predicted by full LQCD \cite{Karsch1994,Kaczmarek,Karsch2001}. 
Therefore we rescale $T_0$ to 190 MeV 
so that the EPNJL model can reproduce $T_{\rm c}=174$~MeV~\cite{Sakai-EPNJL}. 
\begin{table}[h]
\begin{center}
\begin{tabular}{cccc}
\hline \hline
$a_0$ & $a_1$ & $a_2$ & $b_3$
\\
\hline
~~~$3.51$~~ & ~~$-2.47$~~ & ~~$15.2$~~~ & ~~$-1.75$~~\\
\hline \hline
\end{tabular}
\caption{
Summary of the parameter set in the Polyakov-loop potential sector 
determined in Ref.~\cite{Rossner}. 
All parameters are dimensionless. 
}
\label{table-para}
\end{center}
\end{table}

The four-quark vertex originates from the one-gluon exchange between quarks and its higher-order diagrams. 
If the gluon field $A_{\nu}$ has a vacuum expectation value $\langle A_{0} \rangle$ in its time component, 
$A_{\nu}$ is coupled to $\langle A_{0} \rangle$ and then to $\Phi$ through $L$.
Hence the effective four-quark vertex can depend on $\Phi$~\cite{Kondo}.
In this paper, we use the following form for $G(\Phi)$\cite{Sakai-EPNJL}:
\begin{eqnarray}
G(\Phi)=G_{\rm S}[1-\alpha_1\Phi\Phi^*-\alpha_2(\Phi^3+\Phi^{*3})]. 
\label{entanglement-vertex}
\end{eqnarray}
This form preserves the chiral symmetry, the charge conjugation ($C$) symmetry and the extended $\mathbb{Z}_3$ symmetry \cite{Sakai-extendedZ3}.
We take the parameters $(\alpha_1,\alpha_2)=(0.2,0.2)$ to reproduce LQCD data at imaginary $\mu_{\rm B}$ \cite{Sakai-EPNJL}.
It is expected that $\Phi$ dependence of $G(\Phi )$ will be determined in future by the accurate method such as the exact renormalization group method~\cite{Braun,Kondo,Wetterich}. 

Performing the mean-field approximation and 
the path integral over the quark field, 
one can obtain the thermodynamic potential $\Omega$ (per volume): 
\begin{eqnarray}
\frac{\Omega}{V}
&=&
G(\Phi )\sigma^2 + \mathcal{U}
- 2N_{\rm c}\sum_{f={\rm u,d}}
\int_{\Lambda}
\frac{d^3p}{(2\pi )^3}
E_f
\nonumber
\\
&&
-
\frac{2N_{\rm c}}{\beta}\sum_{f={\rm u,d}}
\int
\frac{d^3p}{(2\pi )^3}
\Bigl\{
\ln
\Bigl[
1+3\Phi e^{-\beta (E_f-\mu_f)}
\nonumber\\
&&
+3\Phi^*e^{-2\beta (E_f-\mu_f)}+e^{-3\beta (E_f-\mu_f)}
\Bigr]
\nonumber\\
&&
+
\ln
\Bigl[
1+
3\Phi^*e^{-\beta (E_f+\mu_f)}+3\Phi e^{-2\beta (E_f+\mu_f)}
\nonumber\\
&&
+e^{-3\beta (E_f+\mu_f)}
\Bigr]
\Bigr\}
\label{EPNJL-Omega}
\end{eqnarray}
with
\begin{eqnarray}
E_f = \sqrt{\vec{p}^{~2}+M_f^2}
,~~
M_f = m_0-2G(\Phi )\sigma
,~~
\sigma\equiv\braket{\bar{q}q}.
\end{eqnarray}
The quark-number densities $n_{\rm u}$ and $n_{\rm d}$ are obtained by
\begin{eqnarray}
n_f
=
-
\frac{\partial}{\partial\mu_f}\left(\frac{\Omega}{V}\right)
\end{eqnarray}
for $f={\rm u},{\rm d}$ and the pressure $P$ is defined as $P=-\Omega + \Omega_0$, where $\Omega_0$ is thermodynamic potential at $T=\mu_{\rm u}=\mu_{\rm d}=0$.
The three-dimensional cutoff is introduced for the momentum integration, 
since this model is nonrenormalizable; 
this regularization is denoted by $\int_{\Lambda}$ in Eq. \eqref{EPNJL-Omega}. 
For simplicity, we assume isospin symmetry for ${\rm u}$ and ${\rm d}$ masses: $m_{l} \equiv m_{\rm u}=m_{\rm d}$. 
At $T=0$, the EPNJL model agrees with the NJL model that has three parameters; $G_{\rm S}$, $m_l$, and $\Lambda$. 
One of the typical parameter sets is shown in Table \ref{Table_NJL}~\cite{Kashiwa}. 
These parameters are fitted to empirical values of pion mass and decay constant at vacuum. 
\begin{table}[h]
\begin{center}
\begin{tabular}{ccc}
\hline \hline
~~$m_l({\rm MeV})$~~ & ~~$\Lambda({\rm MeV})$~~ & ~~$G_{\rm S} ({\rm GeV}^{-2})$~~
\\
\hline
$5.5$ & $631.5$ & $5.498$
\\
\hline \hline
\end{tabular}
\caption{
Summary of the parameter set in the NJL sector taken 
from Ref. \cite{Kashiwa}.
\label{Table_NJL}
}
\end{center}
\end{table}

The classical variables $X=\Phi$, ${\Phi}^*$ and $\sigma$
are determined by the stationary conditions 
\begin{equation}
\frac{\partial\Omega}{\partial X}=0. 
\end{equation}
The solutions to the stationary conditions do not give the global minimum of $\Omega$ necessarily. 
They may yield a local minimum or even a maximum. 
We then have checked that the solutions yield the global minimum when  the solutions $X(T, \mu_{\rm u},\mu_{\rm d})$ are inserted into Eq. (\ref{EPNJL-Omega}).
In this work, we employ an approximation $\Phi = \Phi^*$ for numerical simplicity, 
because the approximation is good and hence sufficient for the present analysis\cite{Sakai-signproblem}.
 
Repulsive forces among quarks are crucial to account for the 2$M_{\odot}$ NS observation~\cite{Masuda,Shao}, since they 
harden the EoS of quark matter.
We then introduce the vector-type four-body interaction 
to the EPNJL model \cite{Sakai-vector2008},
\begin{equation}
{\cal L}_{\rm EPNJL}
~\rightarrow~
{\cal L}_{\rm EPNJL}
+G_{\rm V}(\bar{q}\gamma_{\mu}q)^2.
\end{equation} 
The corresponding thermodynamic potential is obtained by the 
replacement,
\begin{eqnarray}
\mu_f
&\rightarrow&
\mu_f - 2G_{\rm V}n_{\rm q},
\\
G(\Phi )\sigma^2
&\rightarrow&
G(\Phi )\sigma^2 - G_{\rm V}n_{\rm q}^2
\end{eqnarray} 
with $n_{\rm q}\equiv\braket{q^{\dagger}q}$. Here, 
$n_{\rm q}$ is determined in a self-consistent manner 
to satisfy the thermodynamic relation,
\begin{equation}
-
\frac{\partial}{\partial \mu_{\rm q}}
\left(\frac{\Omega}{V}\right)
=
n_{\rm q},
\end{equation}
where $\mu_{\rm q}=\mu_{\rm B}/3=(\mu_{\rm u}+\mu_{\rm d})/2$.
The parameter $G_{\rm V}$ is treated as a free parameter in this paper.  
$G_{\rm V}$ dependence of the quark-hadron phase transition 
will be discussed in Sec. \ref{Numericalresults}.
\subsection{Hadron phase}
Now we consider the hadron phase by using the HRG model and its extension. The pressure of  the HRG model is composed of 
meson and baryon parts, 
\begin{equation}
P_{\rm H}
=
P_{\rm M}
+
P_{\rm B}
\end{equation}
where $P_{\rm H}, P_{\rm M}$ and $P_{\rm B}$ are pressures of 
hadronic, mesonic and baryonic matters, respectively.
For the meson part, we use the HRG model with no extension:
\begin{eqnarray}
P_{\rm M}
&=&
\sum_i
d_iT
\int
\frac{d^3p}{(2\pi )^3}
\ln
\left( 1-e^{-\beta E_i}\right)
\\
E_i
&=&
\sqrt{\vec{p}^{~2}+M_i^2},
\end{eqnarray}
where the summation is taken over all meson species 
and $M_i$ and $d_i$ are mass and degeneracy of $i$th meson, respectively.

For  the baryon sector, the volume-exclusion effect \cite{Sakai-EOS,Rischke,Steinheimer} is introduced to reproduce the repulsive nature of the nuclear-matter EoS determined  
from the Ch-EFT 2NF and 3NF and the  heavy-ion collision measurements that will be shown later in Sec. \ref{LOBT}. 

We consider the system of particles having a finite volume $v$, characterized by thermodynamic variables ($T,V,\mu$). 
Following Refs. \cite{Sakai-EOS,Rischke,Steinheimer}, we approximate the system of finite-volume particles by the mimic system of point particles 
with ($T,\tilde{V},\tilde{\mu}$) defined by
\begin{eqnarray}
\tilde{V}&=&V-vN_{\rm B},\\
\tilde{\mu}&=&\mu - vP, 
\end{eqnarray}
where $N_{\rm B}$ is the total baryon number. 
The $P$ and $N_{\rm B}$ should be the same between the original and mimic systems.
The chemical potential $\tilde{\mu}$ of the mimic system is determined to preserve the thermodynamic consistency.
The procedure can be extended to the multi-species system 
composed of proton (p) and neutron (n), 
and
the pressure of the mimic system is obtained by
\begin{eqnarray}
P_{\rm B}
&=&
\frac{2}{\beta}\sum_{i=p,n}
\int\frac{d^3p}{(2\pi )^3}
\Bigl[
\ln
\left(
1+e^{-\beta (E_i-\tilde{\mu}_i)}
\right)
\nonumber\\
&&
+
\ln
\left(
1+e^{-\beta (E_i+\tilde{\mu}_i)}
\right)
\Bigl]
\end{eqnarray}
with $E=\sqrt{p^2+M_i^2}$, $M_{\rm p}=938$~MeV, and $M_{\rm n}=940$ MeV\cite{PDG}. The entropy density ($s$) and 
the number densities ($n_{\rm p},n_{\rm n}$) of  the original system are 
obtained from those of mimic system by
\begin{eqnarray}
s
&=&
\frac{\tilde{s}}{1+v\tilde{n}_{\rm B}},
\\
n_i
&=&
\frac{\tilde{n}_i}{1+v\tilde{n}_{\rm B}},
\end{eqnarray}
with $i={\rm p},{\rm n}$ and $\tilde{n}_{\rm B}=\tilde{n}_{\rm p}+\tilde{n}_{\rm n}$.
\subsection{LOBT calculation with Ch-EFT interactions}
\label{LOBT}
The Brueckner theory is a standard framework to describe nuclear matter starting from realistic 2N interactions. 
The reaction matrix $G$, defined by the $G$-matrix equation
\begin{equation}
 G_{12}=v_{12}+v_{12}\frac{Q}{\omega -(t_1+U_1+t_2+U_2)}G_{12},
\end{equation}
properly deals with short range (high momentum) singularities of the 2N potential $v_{12}$.
The self-consistent determination of the single-particle (s.p.) potential $U$,
\begin{equation}
 \langle i|U|i\rangle \equiv \sum_j^{\rm occupied} \langle ij|G_{12}|ij-ji \rangle
\end{equation}
corresponds to the inclusion of a certain class of higher-order correlations.
In the above expression, $Q$ stands for the Pauli exclusion, $t_i$ is a kinetic energy operator, 
and $\omega$ is a sum of the initial two-nucleon s.p. energies.
The reliability of the lowest-order calculation in the Brueckner theory has been 
demonstrated by the estimation of the smallness of the contribution of higher-order
correlations on the one hand and by the consistency with the results from
other methods such as variational framework\cite{Baldo}.

The Ch-EFT provides a systematic determination of 2NF and 3NF. 
It is prohibitively hard, at present, to do full many-body calculations for infinite matter with including 3NF. 
The effects can be estimated by introducing a density-dependent effective
2N force $v_{12(3)}$ obtained by folding the third nucleon in infinite matter considered:
\begin{eqnarray}
 \langle \bk_1' \sigma_1' \tau_1', \bk_2'\sigma_2'\tau_2'|v_{12(3)}
 |\bk_1 \sigma_1 \tau_1, \bk_2\sigma_2\tau_2\rangle_A \nonumber \\
 =\sum_{\bk_3\sigma_3\tau_3} \langle \bk_1'\sigma_1'\tau_1',
 \bk_2'\sigma_2'\tau_2', \bk_3\sigma_3\tau_3|v_{123} \nonumber \\
 |\bk_1\sigma_1\tau_1,
 \bk_2\sigma_2\tau_2, \bk_3\sigma_3\tau_3\rangle_A, \label{eq:efv}
\end{eqnarray}
where $\sigma$ and $\tau$ stand for the spin and isospin indices,
and two-remaining nucleons are assumed to be in the center-of-mass frame, 
namely $\bk_1'+\bk_2'=\bk_1+\bk_2$. 
The suffix $A$ denotes an antisymmetrized matrix element. 
The $G$-matrix equation is set up for the two-body interaction $v_{12}+\frac{1}{3}v_{12(3)}$. 
The factor $\frac{1}{3}$ is necessary for properly taking into account the combinatorial factor in evaluating the total energy. 
The LOBT $G$-matrix calculation in this approximation turns out to give quantitatively satisfactory description for the fundamental
properties of nucleon many-body systems, namely saturation and strong spin-orbit field: 
the latter is essential for accounting for nuclear shell structure. 
These results were briefly reported in Ref. \cite{MK12}.
Detailed accounts will be given in a separate paper.

In neutron matter, the contact $c_E$ term of the Ch-EFT 3NF vanishes and the $c_D$ term contributes negligibly. 
This means that the 3NF contributions in neutron matter are determined by the parameters that are fixed in the 2NF sector. 
Thus ambiguities concerning the 3NF contributions are minimal with the use of the Ch-EFT, in contrast to past studies in which phenomenological regulations were often applied. 
Because many-body correlation effects are expected not to be large because of the absence
of strong tensor-force correlations in the $^3$E channel, 
the LOBT energies should be reliable in neutron matter.

\begin{figure}[tb]
\begin{center}
 \includegraphics[width=0.45\textwidth]{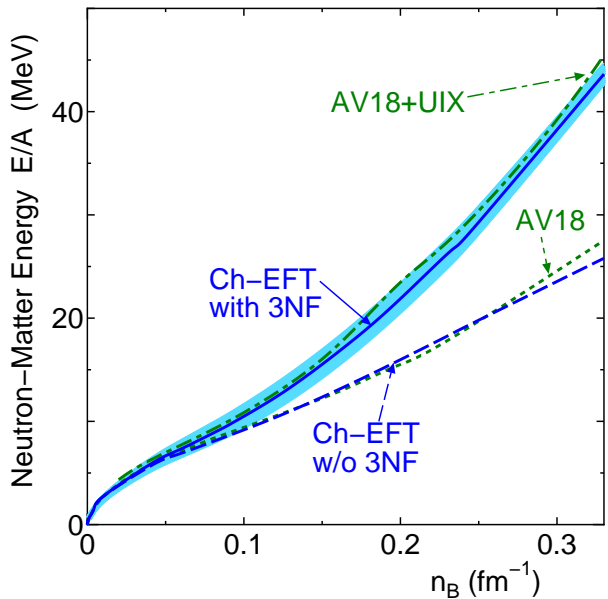}
\vspace{-15pt}
\end{center}
\caption{
Neutron-matter energies as a function of the density $n_{\rm B}$.
The solid and dashed curves are results of the Ch-EFT interactions 
with and without 3NF, respectively. 
The dotted curve shows results of the AV18 2NF \cite{AV18}.
The typical result of the variational method by the Illinois group \cite{APR} is include by a dot-dashed curve, 
in which the Urbana 3NF is used together with the AV18.
}
\label{chiPT}
\end{figure}

Calculated energies of neutron matter with and without 3NF are shown in Fig. \ref{chiPT}, where the cutoff energy $\Lambda_{\rm EFT}$ 
of the Ch-EFT 2NF and 3NF is 550~MeV. 
The solid and dashed curves are results using the Ch-EFT interactions 
with and without 3NF, respectively. 
The energy curve without 3NF is very close to that of the standard modern 2NF, AV18 \cite{AV18}.  
For comparison, energies from the variational calculation by Illinois group \cite{APR} are included, 
which are frequently referred to as the standard EoS for discussing NS  properties although their 3NF is phenomenological to some extent. 
It is interesting that the present prediction based on the Ch-EFT shows good correspondence to those energies.

In the application of the Ch-EFT, an estimation of theoretical uncertainties
due to the uncertainties of the low-energy constants is customarily presented.
As for the neutron-matter EoS, it is instructive to consult the estimation by Kr\"{u}ger {\it et al.} \cite{KTHS}. 
They show, in their Hatree-Fock type calculations that the neutron-matter energy at saturation density is in a range
of $-14 \sim -17$ MeV for the Ch-EFT potential of the J\"ulich group \cite{EHM09} 
with the cutoff parameter of 450/700 MeV from uncertainties of coupling constants 
and cutoff parameters as well as many-body theoretical treatment.
Following this estimation, we add the shaded are to indicate possible uncertainties, 
simply assuming the $\pm 8$ \% of the potential contribution, which is $-18.6$ MeV at saturation density.
%
%
\section{Numerical results}
\label{Numericalresults}
\subsection{Zero temperature}
\label{Zero temperature}
At zero temperature, the present hybrid model becomes simpler. 
Mesons do not contribute to the pressure, and the quark phase is described 
by the NJL model, since the EPNJL model is reduced to the NJL model there.
In this section, we discuss the MR relation of NS, assuming that 
the hadron phase is a neutron-matter system. 

The NJL model for the quark phase is solved under the condition
\begin{equation}
2n_{\rm u} = n_{\rm d},
\end{equation}
and the neutron-number density ($n_{\rm n}$) and its chemical potential ($\mu_{\rm n}$) are given by
\begin{eqnarray}
n_{\rm n}
&=&
\frac{2n_{\rm d}-n_{\rm u}}{3},
\\
\mu_{\rm n}
&=&
\mu_{\rm u}+2\mu_{\rm d}.
\end{eqnarray}

\begin{figure}[tb]
\begin{center}
 \includegraphics[width=0.45\textwidth,bb=30 30 410 300,clip]{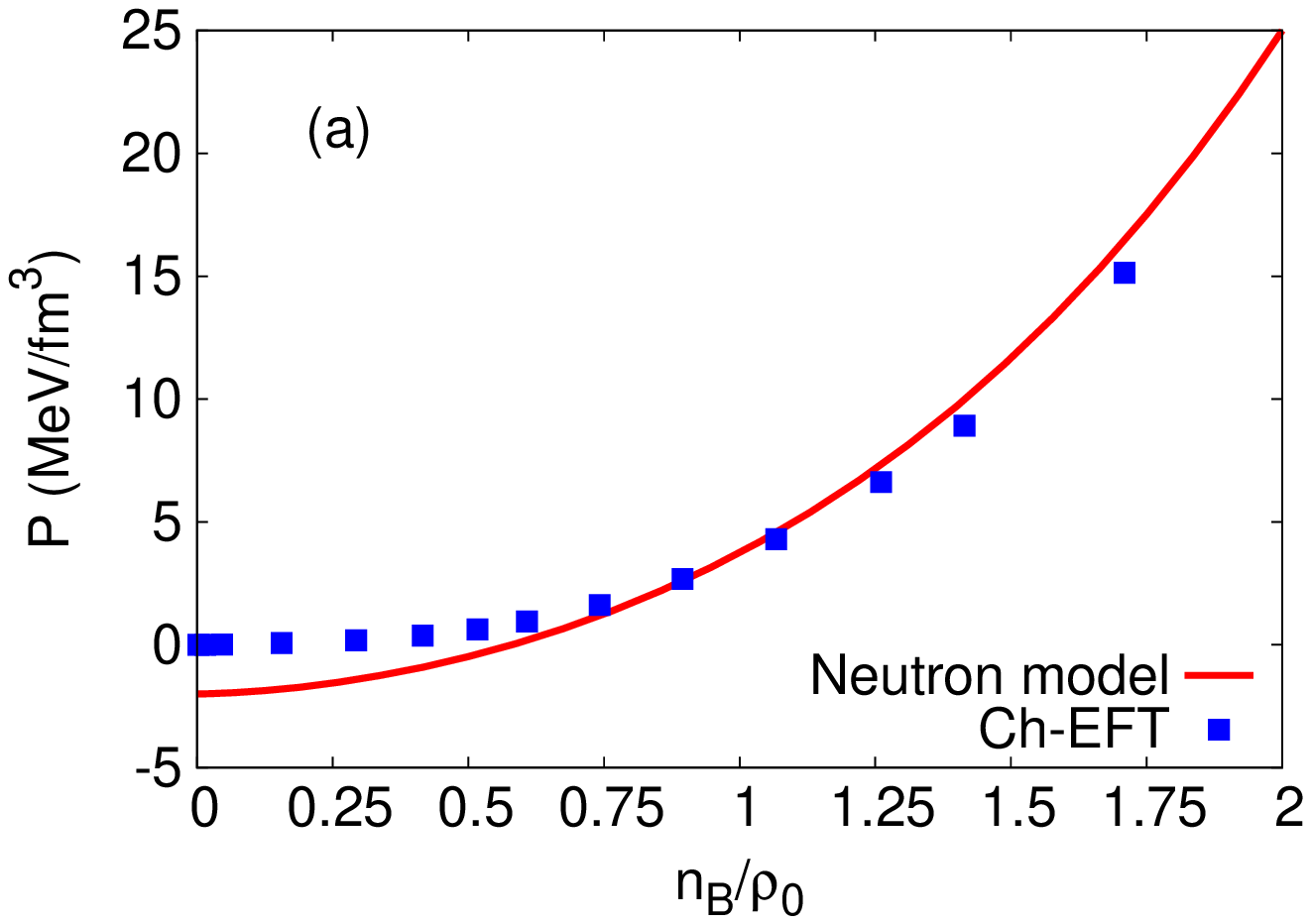}
 \includegraphics[width=0.45\textwidth,bb=30 30 410 300,clip]{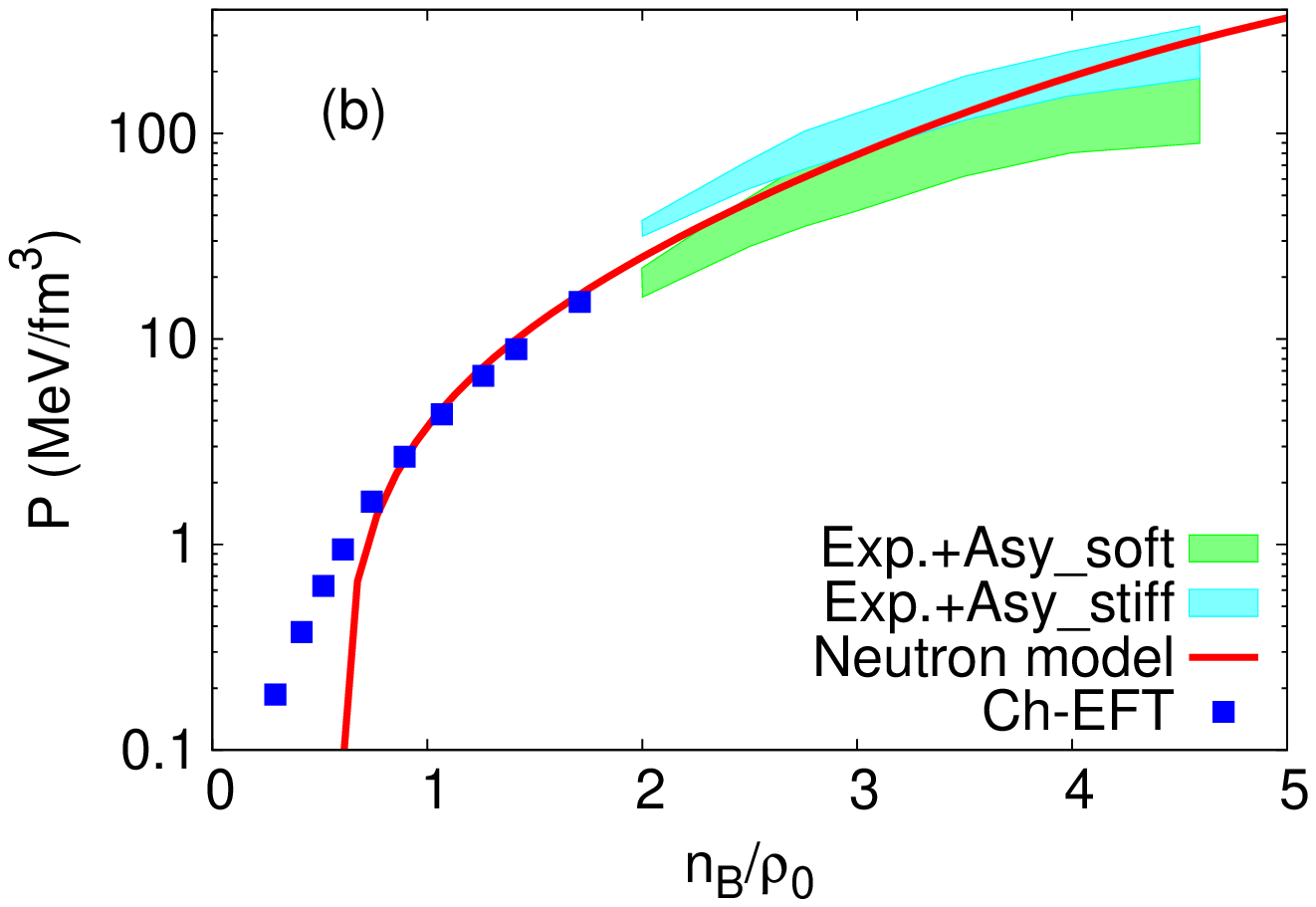}
\vspace{-15pt}
\end{center}
\caption{
Baryon-number density ($n_{\rm B}$) dependence of pressure ($P$) for neutron matter.
$n_{\rm B}$ is normalized by the normal nuclear density $\rho_0=0.17$ (${\rm fm}^{-3}$).
In the panel (b), experimental data is taken from Ref.\cite{Danielewicz}.
}
\label{fitting}
\end{figure}

In the HRG model for the hadron phase, neutrons are assumed to have the exclusion volume $v$ which depends on $\tilde{\mu}_{\rm B}$. 
The dependence is parameterized as 
\begin{eqnarray}
v
&=&
\frac{4}{3}\pi r_{\rm excl}^3,
\\
r_{\rm excl}(\tilde{\mu}_{\rm B})
&=&
r_0+r_1\tilde{\mu}_{\rm B}+r_2\tilde{\mu}_{\rm B}^2.
\end{eqnarray}
Figure \ref{fitting} shows $n_{\rm B}$ dependence of the neutron-matter pressure; 
note that $n_{\rm B}=n_{\rm n}$ in neutron matter and it is normalized by the normal nuclear density $\rho_0=0.17$ (${\rm fm}^{-3}$). 
Closed squares denote the results of LOBT calculations with the Ch-EFT 2NF and 3NF. 
The results are plotted in the region of $n_{\rm B} < 2\rho_0$, since 
the Fermi energy becomes larger than the cutoff energy 
$\Lambda_{\rm ERT}$ beyond $n_{\rm B} = 2\rho_0$. 
As shown in panel (a), the result (solid line) of the HRG model 
with the volume-exclusion effect 
well reproduces the results of LOBT calculations 
at $\rho_0 \lsim n_{\rm B} \lsim 2\rho_0$, 
when  
\begin{eqnarray}
r_0&=&0.50 (\rm fm),\\
r_1&=&0.50 (\rm fm/GeV),\\
r_2&=&- 0.34 (\rm fm/GeV^2) .
\end{eqnarray}
More precisely, the difference between the two results is 
at most $2 ({\rm MeV/fm^3})$, but the deviation is smaller than the theoretical uncertainty of the Ch-EFT EoS estimated in Sec. \ref{LOBT}. 
For $n_{\rm B} < \rho_0$, the agreement of the extended HRG model 
with the Ch-EFT EoS is not perfect, so the Ch-EFT EoS itself is used 
there whenever the MR relation is evaluated. 

In panel (b), the neutron-matter  pressure is plotted at higher $n_{\rm B}$. 
The hatching area shows the empirical EoS \cite{Danielewicz} evaluated from heavy-ion collisions in which the uncertainty coming from the symmetry energy is taken into account. 
The present HRG model is also consistent with this empirical result.
%
%

\begin{figure}[tb]
\begin{center}
 \includegraphics[width=0.45\textwidth,bb=30 30 410 300,clip]{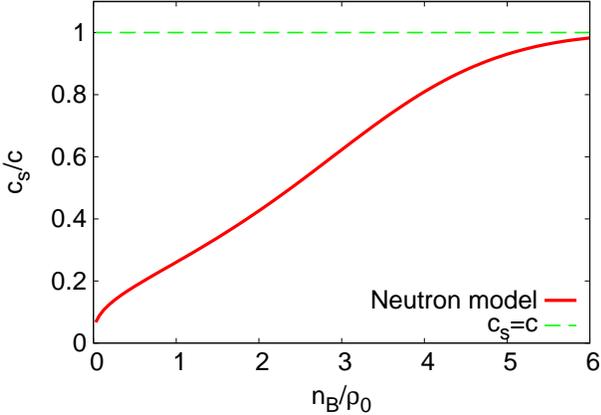}
\vspace{-15pt}
\end{center}
\caption{
Baryon-number density ($n_{\rm B}$) dependence of the speed of sound ($c_{\rm S}$) in neutron matter.
}
\label{sound}
\end{figure}

The speed of sound ($c_{\rm S}$) relative to the speed of light ($c$) is obtained by
\begin{equation}
\frac{c_{\rm S}}{c}
=
\sqrt{\frac{dP}{d\varepsilon}}
\end{equation}
with the energy density $\varepsilon$.
The ratio $c_{\rm S}/c$ should be smaller than 1 to preserve the causality.
As shown in Fig. \ref{sound} that shows $n_{\rm B}$ dependence of $c_{\rm S}/c$, the present HRG model satisfies the causality 
even in the high-density region.
%
%

\begin{figure}[tb]
\begin{center}
 \includegraphics[width=0.45\textwidth,bb=30 30 410 300,clip]{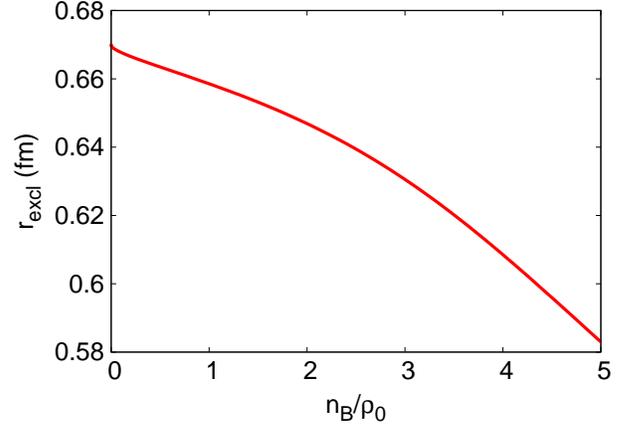}
\vspace{-15pt}
\end{center}
\caption{
Baryon-number density ($n_{\rm B}$) dependence of neutron exclusion radius ($r_{\rm excl}$).
}
\label{ex-radius}
\end{figure}

Figure \ref{ex-radius} shows $n_{\rm B}$ dependence of 
the neutron exclusion radius $r_{\rm excl}$. 
The resulting $r_{\rm excl}$ determined from the Ch-EFT and 
the empirical EoS has weak $n_{\rm B}$ dependence and the value is around 0.6 fm that is not far from the proton charge radius 0.877 fm\cite{PDG}. 
This fact implies that the present model is reasonable as an effective model.  
%
%

The MR relation of NS is obtained by solving the static and spherically symmetric Einstein equation, i.e., the Tolman-Oppenheimer-Volkoff (TOV) equation,
\begin{eqnarray}
\frac{dP}{dr}
&=&
-G_{\rm N}\frac{\varepsilon m}{r^2}
\left(1+\frac{P}{\varepsilon}\right)
\left(1+\frac{4\pi Pr^3}{m}\right)
\left(1-\frac{2G_{\rm N}m}{r}\right)^{-1} ,
\nonumber\\
\frac{dm}{dr}
&=&
4\pi r^2\varepsilon 
\end{eqnarray}
with $G_{\rm N}$ being the gravitational constant \cite{Shapiro}, where 
\begin{eqnarray}
m(r)=\int^r_04\pi r^{\prime 2}\varepsilon (r')dr'
\end{eqnarray}
corresponds to the gravitational mass of the sphere of radius $r$.
The solutions, $m(r)$ and $P(r)$, can be obtained 
by integrating the TOV equations numerically,  
when the EoS, $P=P(\varepsilon)$, is given. 
The integration stops at $r=R$ where $P(R)=0$, and the maximum value  $R$ is the radius of NS and the mass is given by $M=m(R)$.
Here, we adopt the Baym-Pethick-Sutherland (BPS) EoS for the outer crust \cite{Baym}. 
Although, for the inner crust, we should consider the non-uniform structures, 
namely the pasta structures \cite{Maruyama2010}, 
we just connect the outer crust EoS to the Ch-EFT EoS at the subnuclear density smoothly, 
since this simplification does not affect on the MR relation. 
Similarly the Ch-EFT EoS is connected to the HRG-model EoS at  $n_{\rm B}\sim \rho_0$.

\begin{figure}[tb]
\begin{center}
 \includegraphics[width=0.45\textwidth,bb=30 30 410 300,clip]{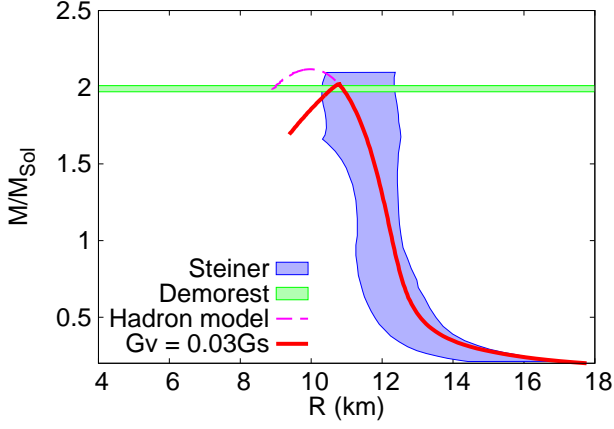}
\vspace{-15pt}
\end{center}
\caption{
The mass-radius relation obtained by the neutron matter with quark-hadron transition.
The two observation data are taken from Ref. \cite{Demorest,Steiner}.
}
\label{MR-full}
\end{figure}

Figure \ref{MR-full} shows the MR relation obtained by the hadron model mentioned above. 
The model result (dashed line) is compared with two observation data.
The first one obtained by A. W. Steiner {\it et al.} is the 
best fitting against various observations on the MR relation \cite{Steiner}.
This is not a strong constraint because of the uncertainty of the analysis particularly on X-ray burst phenomena.
The second one has been obtained by P. B. Demorest {\it et al.} from measurements of pulsar J1614-2230 \cite{Demorest}. 
This yields the lower bound of maximum NS mass, $M=(1.97\pm 0.04)M_{\odot}$ and is a strong constraint.
The present hadron model yields a consistent result with both the observations.
%
%

Next, we consider the quark-hadron transition with the Maxwell construction 
by assuming that the transition is the first-order. 
The transition occurs, when the two phases satisfy the conditions
\begin{eqnarray}
\mu_{\rm u} + 2\mu_{\rm d}&=&\mu_{\rm n},\\
P_{\rm Q}(\mu_{\rm u},\mu_{\rm d})&=&P_{\rm H}(\mu_{\rm n}).
\end{eqnarray}
Here we do not consider the finite-size effects due to the Coulomb interaction and the surface tension \cite {Yasutake-review}. 
We will study these effects on the EoS in the future.

Once the quark phase appears as a consequence of the quark-hadron phase transition, it softens the EoS. 
The quark-matter part of the EoS depends on the strength of  $G_{\rm V}$; more precisely, 
it becomes hard as $G_{\rm V}$ increases. 
Hence, the lower bound of $G_{\rm V}$ is determined from the 2$M_{\odot}$ NS observation. 
The lower bound of such $G_{\rm V}$ is $0.03G_{\rm S}$, as shown below. 
Figure \ref{MR-full} shows the MR-relation determined by the present hybrid model. 
The solid line shows the result of the hybrid model with  
$G_{\rm V}=0.03G_{\rm S}$, while the dashed line represents the result of the hadron model 
that corresponds to the hybrid model with $G_{\rm V}=\infty$.  
Thus the hybrid model is consistent with  the 2$M_{\odot}$ NS observation, when $G_{\rm V} \ge 0.03G_{\rm S}$. 

\subsection{Finite temperature}
In this section, we consider the symmetric matter by setting $\mu_{\rm p}=\mu_{\rm n}=\mu_{\rm B}$ and $\mu_{\rm u}=\mu_{\rm d}=\mu_{\rm B}/3$. Understanding of 
the symmetric matter at finite $T$ is important to elucidate 
early universe or heavy-ion collisions.

\begin{figure}[tb]
\begin{center}
 \includegraphics[width=0.45\textwidth,bb=30 30 410 300,clip]{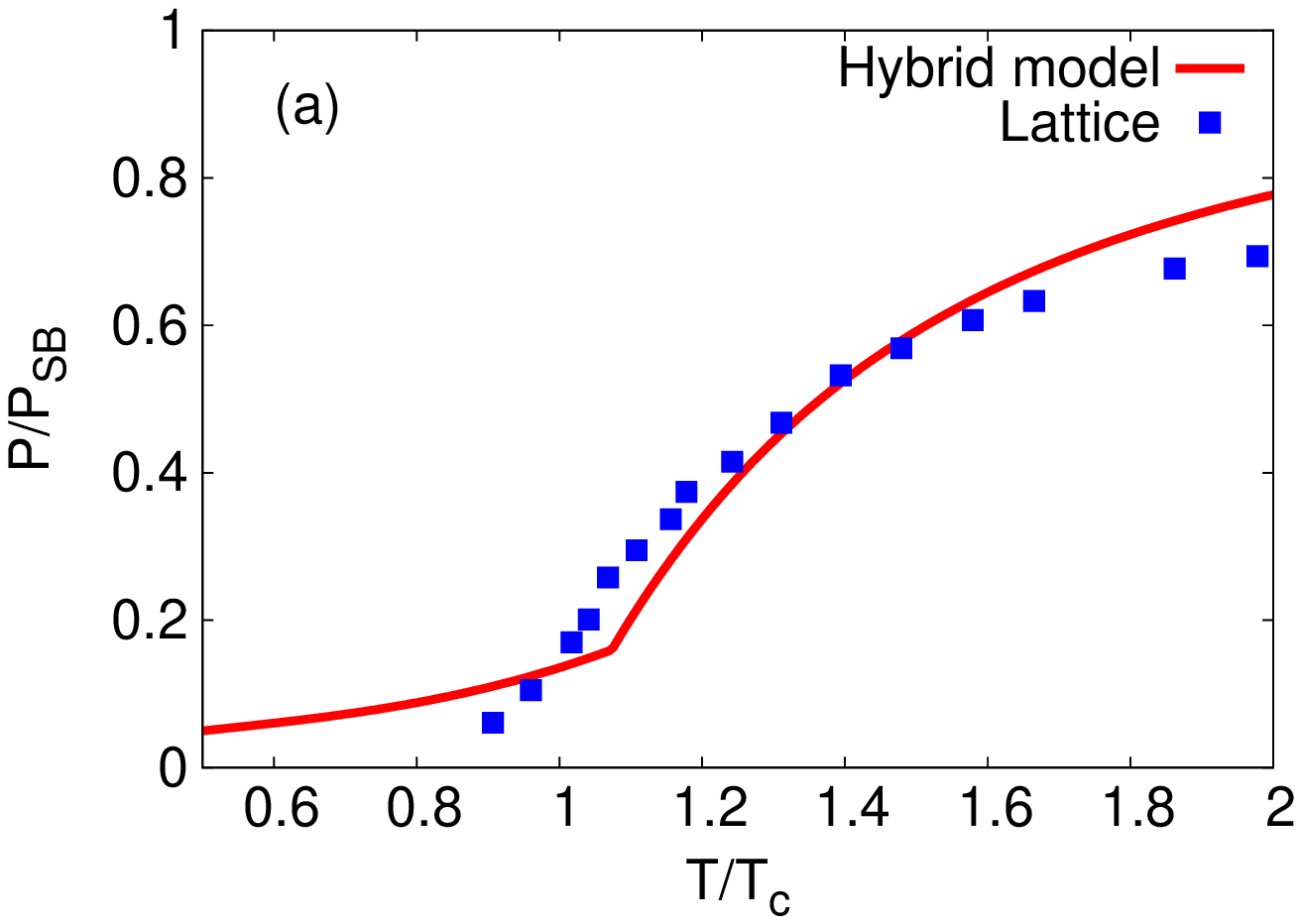}
 \includegraphics[width=0.45\textwidth,bb=30 30 410 300,clip]{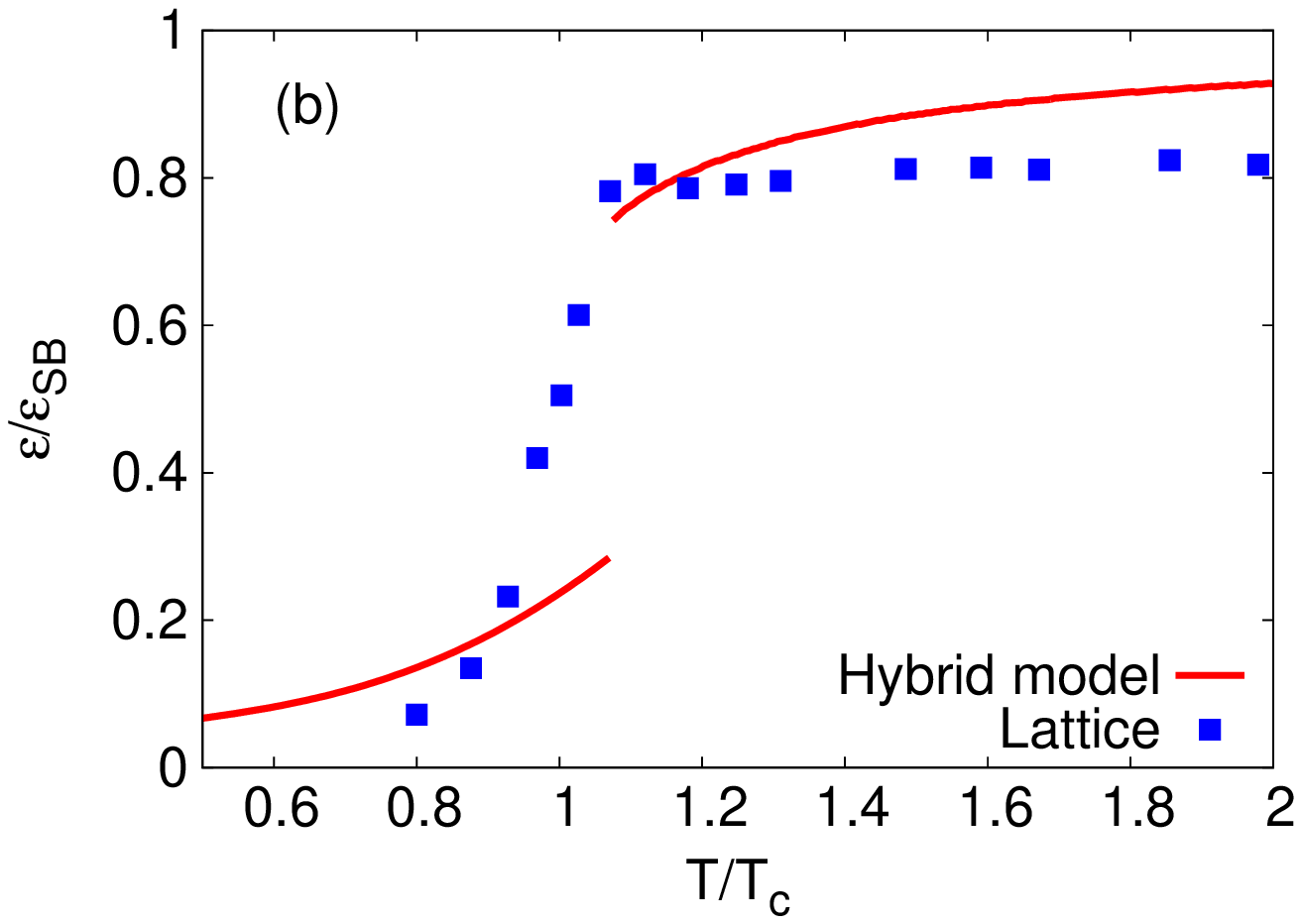}
\vspace{-15pt}
\end{center}
\caption{
T dependence of (a) the pressure and (b) the energy density obtained by the hybrid model.
The result are normalized by their Stefan-Boltzmann limits.
LQCD data is taken from Ref. \cite{AliKhan}.
}
\label{EoS-LQCD}
\end{figure}

Figure \ref{EoS-LQCD} shows $T$ dependence of (a) the pressure and (b) the energy density 
obtained by the hybrid model in comparison with LQCD results \cite{AliKhan}, 
where $T$ is normalized by the deconfinement transition temperature $T_c$.
The deconfinement transition is crossover at $\mu_{\rm B}=0$ in both of LQCD simulations and the EPNJL model. 
The transition temperature defined by the peak of susceptibility
is $T_c=174$~MeV for both the results \cite{Sakai-EPNJL}. 
The hybrid model (solid line) shows the first-order quark-hadron transition, 
whereas the LQCD simulations (closed squares) do the crossover transition. 
Except for the transition temperature $T \approx 1.1T_c$ of the first-order quark-hadron transition, 
the model results almost  reproduce the LQCD results. 
%
%

\begin{figure}[tb]
\begin{center}
 \includegraphics[width=0.45\textwidth,bb=30 30 410 300,clip]{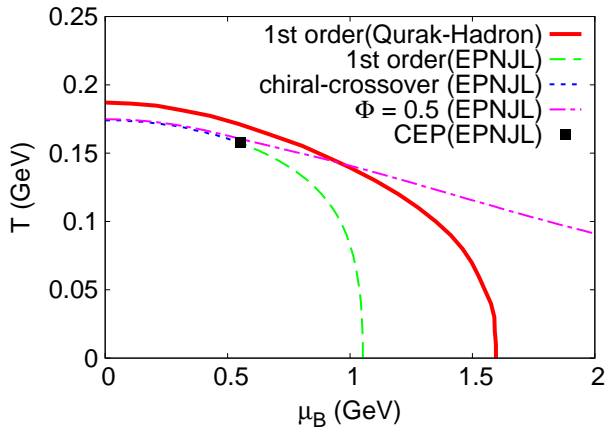}
\vspace{-15pt}
\end{center}
\caption{
Phase diagram in the $\mu_{\rm B}$-$T$ plane.
The solid line represents a quark-hadron transition line given by the hybrid model.
The other lines and symbol are obtained by the EPNJL model.
The dashed (dotted) line correspond to the first-order (crossover) chiral transition line, 
and the dot-dashed line is a contour line corresponds to $\Phi = 0.5$.
The closed square is the critical endpoint (CEP).
}
\label{phase-diagram}
\end{figure}

Figure \ref{phase-diagram} is the phase diagram in the $\mu_{\rm B}$-$T$ plane.
The thick solid line is the quark-hadron transition line obtained by the hybrid model with $G_{\rm V}=0.03G_{\rm S}$. The transition is the first order everywhere. In this sense, this is an approximate result at least at $\mu_{\rm B}/T < 1$, 
since LQCD simulations show that the deconfinement (quark-hadron) transition is crossover there. 
As an important result, the first-order quark-hadron transition line 
is close to the crossover deconfinement transition line (dot-dashed line) obtained by the EPNJL model at $\mu_{\rm B}/T < 1$, 
where the deconfinement transition line is simply defined as 
a line satisfying $\Phi = 0.5$. 
Noting that the EPNJL model well simulates LQCD results at $\mu_{\rm B}/T < 1$, one can see that the present hybrid model is a rather good effective model even at small  $\mu_{\rm B}/T$. 
The dashed and dotted lines correspond to the first-order and crossover 
chiral transition lines, whereas the closed square is the critical endpoint (CEP) of the chiral transition. 

\begin{figure}[tb]
\begin{center}
 \includegraphics[width=0.45\textwidth,bb=30 30 410 300,clip]{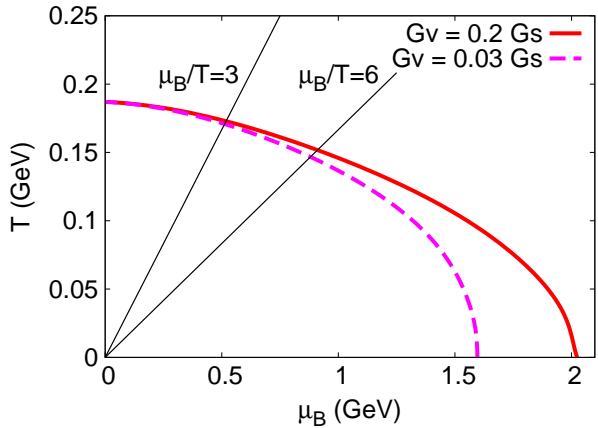}
\vspace{-15pt}
\end{center}
\caption{
Phase diagram in the $\mu_{\rm B}$-$T$ plane.
The dashed line is the result of the hybrid model with $G_{\rm V}=0.03 G_{\rm S}$; the line corresponds to the thick solid line in Fig. \ref{phase-diagram}. 
The thick-solid line corresponds to the case of $G_{\rm V}=0.2 G_{\rm S}$.
Two thin-solid lines mean lines of $\mu_{\rm B}/T=3$ and $6$, respectively.
}
\label{phase-diagram2}
\end{figure}

As already mentioned in Sec. \ref{Zero temperature}, the present hybrid model is consistent with the NS observations at $T=0$, 
when $G_{\rm V} \ge 0.03G_{\rm S}$.
In the hybrid model with $G_{\rm V}=0.03G_{\rm S}$, the critical baryon-number chemical potential $\mu_{\rm B}^{\rm (c)}$ 
of the first-order quark-hadron transition at $T=0$ is $1.6$~GeV, as shown in Fig. \ref{phase-diagram}. 
This is the lower bound of $\mu_{\rm B}^{\rm (c)}$, since $G_{\rm V}$ can vary from $0.03G_{\rm S}$ to $\infty$; 
actually, $\mu_{\rm B}^{\rm (c)}$ is shifted to higher $\mu_{\rm B}$ as $G_{\rm V}$ increases, 
as shown later in Fig. \ref{phase-diagram2}. 
This is the primary result of the present work. 
In the EPNJL model, meanwhile, the critical baryon-number chemical potential of the chiral transition at $T=0$ is 1~GeV. 
The point belongs to the hadron phase in the hybrid model. 
Thus, we do not have any conclusive result on the chiral transition at $T=0$.
This is an important problem to be solved in future. 
%
%
%
%

In principle one can determine the strength of $G_{\rm V}$ from LQCD simulations present at $\mu_{\rm B} /T<3$, but 
in practice the strength thus determined has large ambiguity\cite{Bratovic, Lourenco}.
Figure \ref{phase-diagram2} shows the phase diagram in the $\mu_{\rm B}$-$T$ plane predicted by the hybrid model with 
different values of  $G_{\rm V}$.
The dashed and solid lines correspond to the cases of $G_{\rm V}=0.03 G_{\rm S}$ and $0.2 G_{\rm S}$, respectively. 
The phase transition line is insensitive to the variance of  $G_{\rm V}$ at  $\mu_{\rm B} /T<3$, but rather 
sensitive at $\mu_{\rm B}/T \approx 6$. Thus the physics at $\mu_{\rm B}/T \approx 6$ is strongly related to the NS physics 
at $\mu_{\rm B}/T = \infty$.
If the quark-hadron transition line at $\mu_{\rm B}/T \approx 6$ is determined by LQCD simulations 
or heavy-ion collision experiments, it will also determine $\mu_{\rm B}^{\rm (c)}$ more strictly. 

\section{Summary}
\label{Summary}
We have studied the QCD phase diagram in the whole region from  $\mu_{\rm B}/T=0$ to infinity, 
constructing the quark-hadron hybrid model 
that is consistent with LQCD results at $\mu_{\rm B}/T=0$ and 
at $\mu_{\rm B}/T=\infty$ with NS observations and 
the neutron-matter EoS evaluated from the Ch-EFT 2NF and 3NF and the heavy-ion collision measurements. 
The EoS provided by the model preserves the causality even at high $n_{\rm B}$. 
At $n_{\rm B} < 2\rho_0$ the baryon part of the EoS agrees with the neutron-matter EoS constructed from 
the Ch-EFT 2NF and 3NF with the lowest-order Brueckner theory (LOBT). 
The Ch-EFT provides a systematic framework of constructing 2NF and 3NF, and the 3NF yields a significant effect on the EoS 
at $n_{\rm B} > \rho_0$. 
In this sense, the use of the Ch-EFT, which respects symmetries of QCD, is inevitable to construct the neutron-matter EoS with no ambiguity.

We have determined the lower bound of the critical chemical potential of the quark-hadron transition at $T=0$: 
\begin{equation}
\mu_{\rm B}^{\rm (c)}\sim 1.6~{\rm GeV}. 
\end{equation}
This is the primary result of this work. 
In the NJL model, the first-order chiral transition occurs at  $\mu_{\rm B}^{\rm (c)}=1$~GeV, when $T=0$. 
The point is located in the hadron phase in the hybrid model. 
Thus, the critical chemical potential of the chiral transition at $T=0$ is unknown. 
In this sense, the NJL model is not good enough at $T=0$. 
It is then highly required to introduce baryon degrees of freedom in the effective model. 

We have also shown the interplay between the heavy-ion collision physics at $\mu_{\rm B}/T \approx 6$ and the NS physics at $\mu_{\rm B}/T = \infty$. 
If the vector coupling $G_{\rm V}$ is determined at $\mu_{\rm B}/T \approx 6$ from heavy-ion collision measurements, 
the information determines the critical chemical potential of the quark-hadron transition at $T=0$ and hence properties of NS in the inner core. 
This fact strongly suggests that these two regions should be studied simultaneously.    
\begin{acknowledgments}
The authors thank A. Nakamura and K. Nagano for useful discussions. 
T.S. is supported by JSPS KAKENHI Grant No. 23-2790.
N.Y. is supported by JSPS KAKENHI Grant No. 2510-5510.
M.K. is supported by a Grant-in-Aid for Scientific Research (C) from
the Japan Society for the Promotion of Science (Grant Nos. 22540288 and 25400266).
\end{acknowledgments}

\end{document}